\documentclass[review,number,sort&compress, 5p]{elsarticle}

\usepackage{natbib}

\usepackage{amsmath}
\usepackage{amssymb}
\usepackage{graphicx}
\usepackage{multicol}
\journal{Nuclear Instruments and Methods in Physics Research Section A}

\begin{document}

\begin{frontmatter}

\title{Electron beam test of key elements of the laser-based calibration system for the muon $g$\,$-$\,$2$  experiment}

\author[1,3]{A.~Anastasi\corref{cor2}}
\author[16]{A.~Basti}
\author[16]{F.~Bedeschi}
\author[16]{M.~Bartolini}
\author[4,10]{G.~Cantatore}
\author[4,12]{D.~Cauz}
\author[1]{G.~Corradi}
\author[1,20]{S.~Dabagov}
\author[6]{G.~Di Sciascio}
\author[13,5]{R.~Di Stefano}
\author[12]{A.~ Driutti}
\author[11]{O.~Escalante}
\author[1,2]{ C.~Ferrari}
\author[15]{A.T.~Fienberg}
\author[1,2]{A.~Fioretti}
\author[1,2]{C.~Gabbanini}
\author[18,19]{A.~ Gioiosa}
\author[1]{D.~Hampai}
\author[15]{D.W.~Hertzog}
\author[5,11]{M.~Iacovacci}
\author[4,14]{M.~Karuza}
\author[15]{J.~Kaspar}
\author[1]{A.~Liedl}
\author[16,17]{A.~Lusiani}
\author[13,5]{F.~Marignetti}
\author[5]{S.~Mastroianni}
\author[6]{D.~Moricciani}
\author[4,12]{G.~Pauletta}
\author[18,19]{G.M.~Piacentino}
\author[6]{N.~Raha}
\author[1]{E.~ Rossi}
\author[4]{L.~Santi}
\author[1]{G.~Venanzoni}
\address[1]{Laboratori Nazionali Frascati dell' INFN, Via E. Fermi 40, 00044 Frascati, Italy }
\address[2]{Istituto Nazionale di Ottica del C.N.R., UOS Pisa, via Moruzzi 1, 56124, Pisa, Italy }
\address[3]{Dipartimento MIFT, Universit\`a di Messina, Messina, Italy}
\address[4]{INFN, Sezione di Trieste e G.C. di Udine, Italy}
\address[5]{INFN, Sezione di Napoli, Italy}
\address[6]{INFN, Sezione di Roma Tor Vergata, Roma, Italy}
\address[10]{Universit\`a di Trieste, Trieste, Italy}
\address[11]{Universit\`a di Napoli, Napoli, Italy}
\address[12]{Universit\`a di Udine, Udine, Italy}
\address[13]{Universit\`a di Cassino, Cassino, Italy}
\address[14]{University of Rijeka, Rijeka, Croatia}
\address[15]{University of Washington, Box 351560, Seattle, WA 98195, USA}
\address[16]{INFN, Sezione di Pisa, Italy}
\address[17]{Scuola Normale Superiore, Pisa, Italy}
\address[18]{INFN, Sezione di Lecce, Italy}
\address[19]{Universit\`a del Molise, Pesche, Italy}
\address[20]{Lebedev Physical Institute and NRNU MEPhI, Moscow, Russia}
\cortext[cor2]{Corresponding author: antonioanastasi89@gmail.com}


\begin{abstract}
\noindent \normalsize
We report the test of many of the key elements of the laser-based calibration system for muon $g$\,$-$\,$2$ experiment E989 at Fermilab. The test was performed at the Laboratori Nazionali di Frascati's Beam Test Facility using a 450 MeV electron beam impinging on a small subset of the final $g$\,$-$\,$2$ lead-fluoride crystal calorimeter system. The calibration system was configured as planned for the E989 experiment and uses the same type of laser and most of the final optical elements. We show results regarding the calorimeter's response calibration, the maximum equivalent electron energy which can be provided by the laser and the stability of the calibration system components.
\end{abstract}


\end{frontmatter}



\section{Introduction}
\label{intro}

The  muon $g$\,$-$\,$2$ experiment at Fermilab (E989)  plans to measure the muon anomaly with a total uncertainty of $1.6 \times 10^{-10}$ (0.14 ppm),  which includes a 0.10 ppm statistical error and about 0.07 ppm systematic uncertainties both on the muon anomalous precession angular velocity $\omega_a$ and on the magnetic field measurement with the proton Larmor
precession angular velocity $\omega_p$  ~\cite{carey09}. The new experiment efficiently uses the unique properties of the Fermilab beam complex to produce the necessary flux of 3.1 GeV muons, which will be injected and stored in the (relocated) muon storage ring. To achieve a statistical uncertainty of 0.14 ppm, the total data set must contain more than $1.8  \times 10^{11}$  positrons with energy greater than 1.8 GeV.

The energies of these positrons will be measured by 24 crystal-based calorimeters distributed around the ring and the response of each of the 1296 channels  must be calibrated and monitored to keep uncertainties due to gain variations at the sub-per mil level in the time interval corresponding to one beam fill (700 $\mu$s).  On longer timescales, the goal is to keep systematic contributions due to gain fluctuations at the sub-percent level.

The full E989 calibration system will employ a suite of six identical diode laser light sources, all fired by a common driver.  The light pulses from the lasers are simultaneously injected into the 1296 calorimeter crystals which are viewed by silicon photomultiplier (SiPM) photo-detectors ~\cite{grange}. The laser light pulses closely resemble the Cherenkov light pulses from positron showers in the crystals. Since the laser light pulses originate from a common source they will provide a reliable reference for the time of detection and for positron energy measurement. They can also be used to equalize the response of different calorimeter elements.
The laser calibration system will monitor the intensity of the common
light source and the stability of the light distribution system to the crystals, which may be affected by laser beam pointing
fluctuations, mechanical vibrations or the aging of the transmission elements.

In this paper we report on a test of the key elements of the full E989 calibration system that were employed during a beam test using a subset of the calorimeter.  The Beam Test Facility (BTF) of the Laboratori Nazionali di Frascati \cite{BTF} was used to provide a monoenergetic 450 MeV electron beam, which established the absolute energy scale.  The objectives of the measurements included:

\begin{itemize}
\item Test of the complete calibration system chain (from the control board to the calorimeter).
\item Calibration of the equivalent luminous energy of the laser by comparing the intensity of the laser calibration signals to that produced in the crystals by the electron beam.

\end{itemize}
In the following sections the experimental setup and results will be described.

\section{Experimental setup}
The Frascati BTF provides a highly collimated  electron beam with a 50 Hz repetition rate. We chose to run at very low beam intensities. Most of the runs are taken with an average multiplicity of about one electron per pulse. Higher multiplicities, of up to three electrons per pulse, were also used. The electron beam arrives in the test area with a transverse dimension of about 250 $\mu$m  and a mean position stable in time. The electron energy is selectable from 100 to 500 MeV with a resolution of 1 percent. The calorimeter was positioned on a movable table in order to match the position of the electron beam. The calorimeter consists of a small scale array
of the $PbF_2$ crystals that will be used for the $g$\,$-$\,$2$ experiment. Each of the calorimeters that will instrument Muon $g$\,$-$\,$2$ will be an array of 54 crystals~\cite{carey09}, while the calorimeter used for this test was composed of only five elements (crystals and photo-detectors) arranged in a cross-like configuration with four additional Plexiglass mock crystals so as to create a $3   \times 3$ array.  The sensitive elements are $2.5  \times 2.5  \times14$ cm$^{3}$  high-quality PbF$_2$ crystals~\cite{fienberg15}. Four of them were wrapped in  black absorbing Tedlar, while the fifth was wrapped in reflective white Millipore paper.
A 16-channel Hamamatsu SiPM was glued to the rear face of each crystal~\cite{sipm}. 
\begin{figure}[h]
\includegraphics[width= 0.45\textwidth]{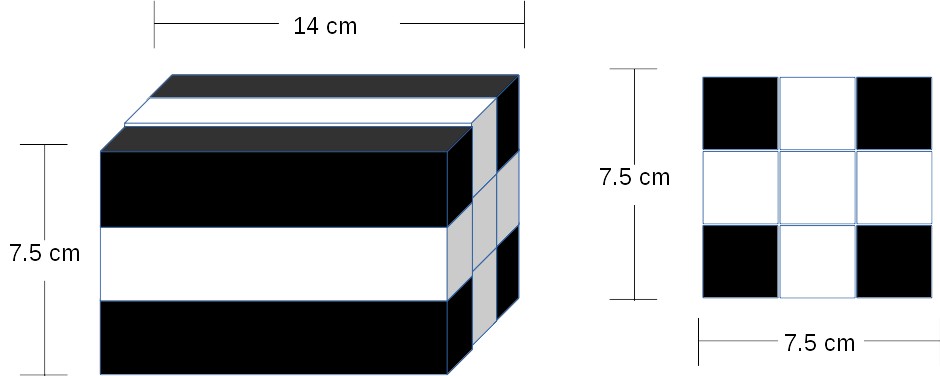}
\caption{Sketch of the crystal configuration of the test calorimeter. Shown in white are the $PbF_2$ crystals and in black the Plexiglas dummies.). \label{calo2}}
\end{figure}
The five SiPMs detected  both the Cherenkov light generated by the beam electrons and the calibration light pulses. Laser calibration pulses were guided to the front face of each calorimeter element by means of optical fibers, each ending on a reflective right-angle prism that injected the light in a direction parallel to the crystal axis. The prisms and each incoming fiber are held by a Delrin panel that is positioned in front of the calorimeter. This panel mocked up all of the final design features related to the panel for a full 54-crystal calorimeter.
\begin{figure}[!h]
\includegraphics[width=.45 \textwidth]{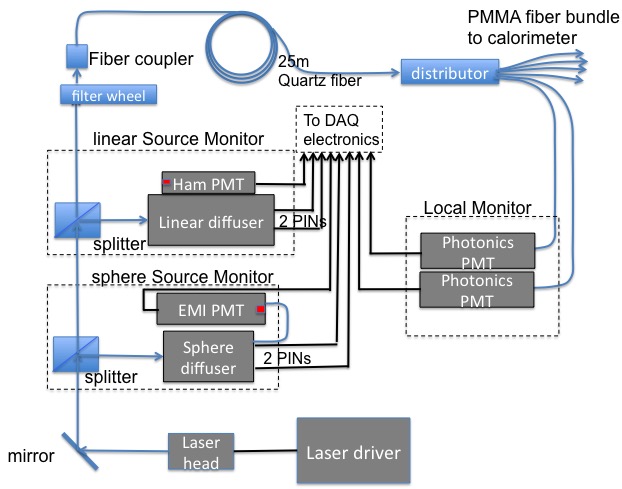}
\caption{Schematic representation  of the experimental layout.}
\label{scheme}
\end{figure}

Each SiPM is connected to the digitizer by a custom PIN-to-MCX signal cable and to a custom breakout board that provides the Bias Voltage by an HDMI cable.
The breakout board is also linked to a BeagleBone microprocessor~\cite{beagle} which is used to run control procedures and set the parameters of the SiPM frontend electronics, e.g. to set gain values for each single SiPM, and to read out its temperature. It communicates with the SiPMs through the HDMI cables of the breakout board. A fan system provided cooling of the SiPMs and their front end electronics.

\subsection{Laser system
\label{ssec:xxx}}

The experiment tested the full distribution chain that will be used to send light to all 1296 channels of  the  muon $g$\,$-$\,$2$  experiment. The setup is illustrated schematically in Figure~\ref{scheme}. The light source is a pulsed laser manufactored by PicoQuant (LDH-P-C-405M), which has a maximum pulse energy of 1 nJ, a  pulse width of about 700 ps at a wavelength of 405 $\pm$10 nm, with a maximum repetition rate of 40 MHz. A laser control board provides the trigger to the laser driver. It also sends out the 
time  reference  signal  for  reset and  synchronization during the
initialization of the detector and the electronics. Moreover, in order to
simulate the physics events in a calibration run without beam to test the
detector and DAQ response, hit patterns can be generated randomly
according to 
an exponential law as expected from muon decay. The system
has been realized by using an hybrid platform with  FPGA  board  and 
ARM-based  processor for configuration and monitoring.

\subsubsection{Laser distribution system
\label{ssec:xxx1}}

Thirty percent of the primary laser beam is re-directed using a splitter cube to the monitoring system that will be described in the next section. The un-deflected laser beam is coupled into a 400 $\mu$m diameter fused silica fiber, with an attenuation of 30 dB/km at 400 nm. This fiber is 25 m long in order to simulate the running conditions of the E989 experiment. The fiber output is recollimated and transmitted through an engineered diffuser produced  by RPC Photonics (mod. ED1-S20), consisting of structured microlens arrays that transform a Gaussian input beam into a flat top one~\cite{anastasi15}. A fiber bundle made of 1 mm-diameter PMMA fibers is positioned about 4 cm from the diffuser. Five of the fibers, each 3 m long, are  connected to the light distribution panel as described in the previous section, two other fibers are connected to two separate photomultipliers (PMTs) which are part of the Local Monitor (see next section).  A motorized neutral-density filter wheel placed before the silica fiber is used to change the intensity of the  laser pulse reaching the calorimeter.


\subsubsection{Monitoring system
\label{ssec:yyy}}

The monitoring system consists of a Source Monitor and a Local Monitor. The Source Monitor (SM) directly measures the laser intensity at the source using thirty percent of the laser light delivered to it by a beam splitter. The SM is designed to reach the required statistical precision rapidly while minimizing sensitivity to extraneous fluctuations due to mechanical vibrations, its own gain, external electronic noise, and to variations in beam pointing and temperature. This design should enable correction of shot-to-shot fluctuations at the per mil level and of variations in the average intensity at the required (0.01\%) precision in about 100 shots. In addition, the light response of the SM is monitored over a longer period by incorporating an absolute reference light source which allows monitoring of average variations over a longer period (hours).
\\
Two different designs of the SM were tested. In both cases, all elements, including the optical splitter, are incorporated in a rigid mechanical structure with a large thermal inertia and good electrical shielding. The designs differ with respect to the method employed to eliminate beam-pointing fluctuations. In the first case the  light from the splitter is mixed by a combination of an engineered diffuser and a reflective mixing chamber while, in the second case, this combination is replaced by an integrating sphere.  In both cases, the mixed light is viewed by a redundant system of 2 large-area PIN diodes (PiDs) and a PMT via a wavelength shifter (WLS). The PMT also views the light signal generated by an $^{241}$ Am radioactive source coupled to a NaI crystal. The PMT therefore detects both the laser light signal and the reference signal provided by the $\alpha$ particles emitted by the weak Americium source at a rate of few Hz. The reference signal serves to correct for possible instabilities in the PMT gain and, since both the PMT and the PiDs see the same laser signal, it serves to control  the stability of the PiDs in a
time interval sufficient to accumulate the required source statistics. 
\\
The PiDs are inherently stable, unity gain, high speed devices that operate at low bias. Their signal must be amplified electronically in this calibration system because of the low available laser intensities. The electronic amplification introduces a
potential instability on the laser intensity monitoring, which can be  monitored by the absolute reference signal
provided by the $^{241}Am$ decays in the PMT.  The frontend electronics used for this test were a simplified version of the devices that are being designed for the $g$\,$-$\,$2$ experiment. 
\\

\begin{figure}[h] \center
\includegraphics[width= 0.45\textwidth]{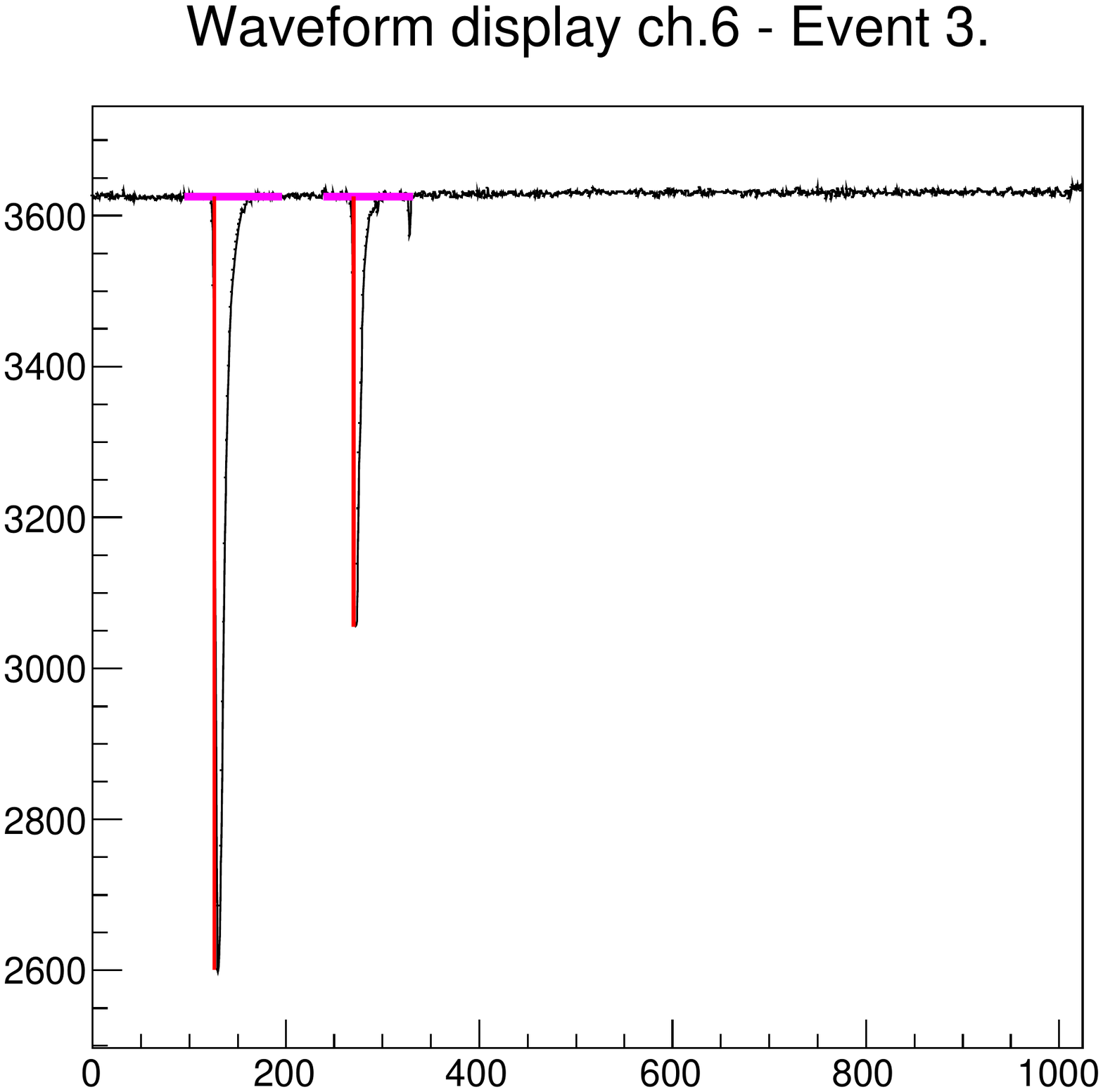}
\caption{Display of a typical LM event. The first signal is direct from the SM while the second is the return signal from the calorimeter. Horizontal scale is nanoseconds.}
\label{LM}
\end{figure}

The SMs also furnish reference signals to the Local Monitors (LM) via optical fibers.  In this configuration the LM receives a fraction below one percent of the light pulse sent to the SM. The LM compares the light intensity on the crystals, at the end of the distribution chain, with that of the light source. In this way it allows to monitor and correct instabilities introduced by the light distributions components.
The LM is a redundant system composed of two Photonics PMTs. Each PMT receives light signals from two fibers: the first fiber comes from the SM and provides the source reference signal while the second fiber comes directly from the bundle which distributes light to the calorimeter crystals. The two pulses are well separated in time by 120 ns, as shown in Fig.~\ref{LM}. The ratio of intensities of the second pulse to the first is a direct measurement of the stability of the distribution chain. The short time span between the signals minimizes the possibility of PMT gain drifts between the two signals.

\subsubsection{Acquisition system
\label{ssec:zz}}
\begin{figure}[ht] \center
\includegraphics[width= 0.45\textwidth]{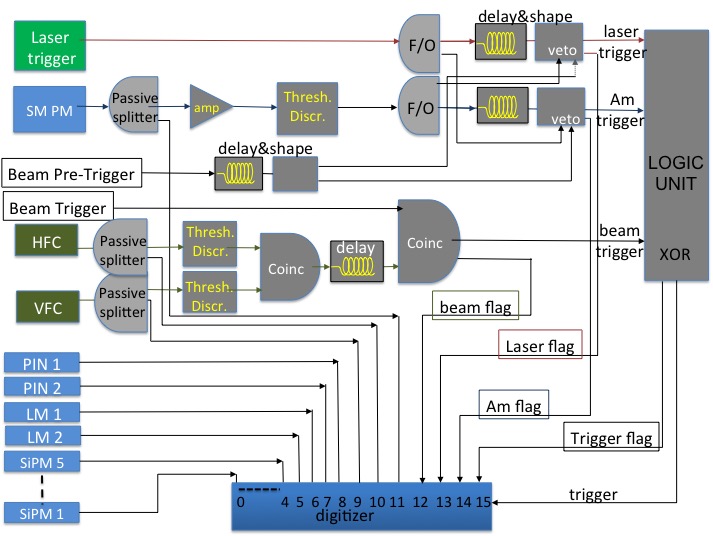}
\caption{Schematic representation of the trigger logic and data acquisition system.}
\label{trigger}
\end{figure}

Two CAEN DT5742 16-channel digitizers sampling at 5 GS/s instrumented the 18 active channels in the test beam. 
Four separate triggers could initiate digitization and readout by the DAQ: a beam trigger, a laser trigger, and an Americium trigger from each of the two SM being tested. 
Fig.\ref{trigger} illustrates the trigger configuration.
Data from temperature sensors, including  ambient, SiPM and electronic board temperatures, were also acquired.

\section{Results}

\subsection{Calibration of the light yield to electron energy
\label{ssec:x}}
 We ran a laser calibration procedure after every configuration change and before all runs with electrons. Besides testing the functionality of all system components, this procedure determines the proportionality constant relating the digitizer output to the number of SiPM pixels fired.

\begin{figure}[ht] \center
\includegraphics[width= 0.45\textwidth]{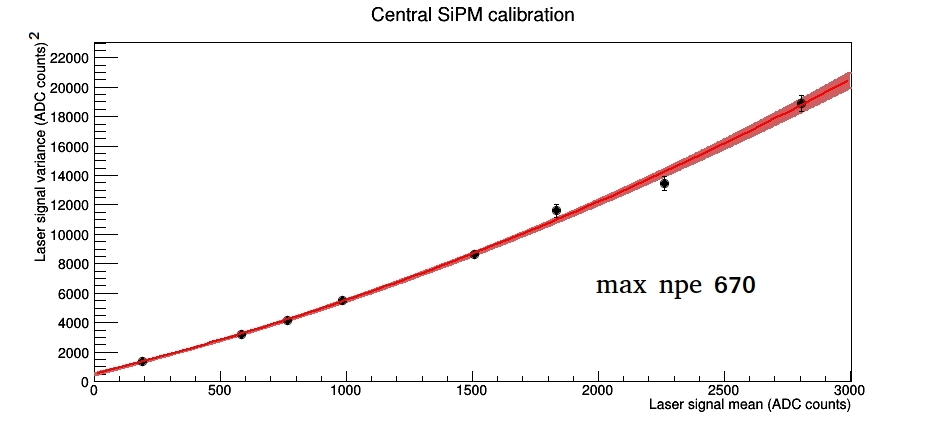}
\caption{An example of a SiPM calibration. Solid points represent measurements with different attenuations of the laser intensity. The fitted curve corresponds to eq.1. The red band reflects the uncertainties in the fit parameters.}
\label{calib}
\end{figure}

These calibrations consist of a series of runs taken with different settings of the filter wheel. We  typically take five thousand laser pulses per run at a frequency of 50 Hz, so it takes only a few minutes per setting. For each setting we measure the mean $\mu$  and the standard deviation $\sigma$  of the distribution of each of the five SiPM signals. In general the signal, $L$,  observed by each SiPM in ADC counts is given by $L=k\nu$, where $\nu$ is the number of pixels fired. The uncertainty in $L$ is due to three main contributions: 1) the electronic noise, $\sigma_N$, 2) the Poisson statistics in the number of fired pixels, $\sigma_P=k\sqrt{\nu}$, and 3) the intrinsic laser pulse fluctuations $\sigma_L=\alpha k\nu$.  The average relative laser intensity variation $\alpha$ has been measured to be less than 1\%.  Other contributions  proportional to $L$ arise from the statistical variation in the number of photons incident on the SiPM photocathodes and from fluctuations in the amplification mechanism.  Based on this model and assuming statistical independence of the sources of fluctuations, the dependence of $\sigma^2$, as a function of the measured light intensity, is given by:
\begin{equation}
 \sigma^2 = \sigma_N^2+kL+\beta L^2 
\end{equation}
where $\beta$ includes all contributions  proportional to $L$. A typical fit of the variance versus signal strength is shown in Fig.~\ref{calib}.  We measure between 600 and 800 fired pixels, depending on SiPM, bias voltage and temperature, when no filtering is applied by the filter wheel. This is about 1\% of the 57600 pixels contained in each SiPM \cite{fienberg15}, where saturation corrections are expected to be in the order of 0.5\% and have negligible impact on these calibration results.

During the runs with electron beam we also pulse the laser at a comparable frequency of 50 Hz.  This provides a reference to relate the laser intensity to the electron beam energy and also allows to calibrate slow variations of the SiPM response during the runs with beam. The calorimeter response to the beam is taken to be the sum of all SiPMs normalized to the response of the central one after correcting for the laser calibrations. This is a reasonable choice given that the beam is strongly focused on the central calorimeter crystal, which collects about 90\% of the electron energy.

\begin{figure}[ht] \center
\includegraphics[width= 0.45\textwidth]{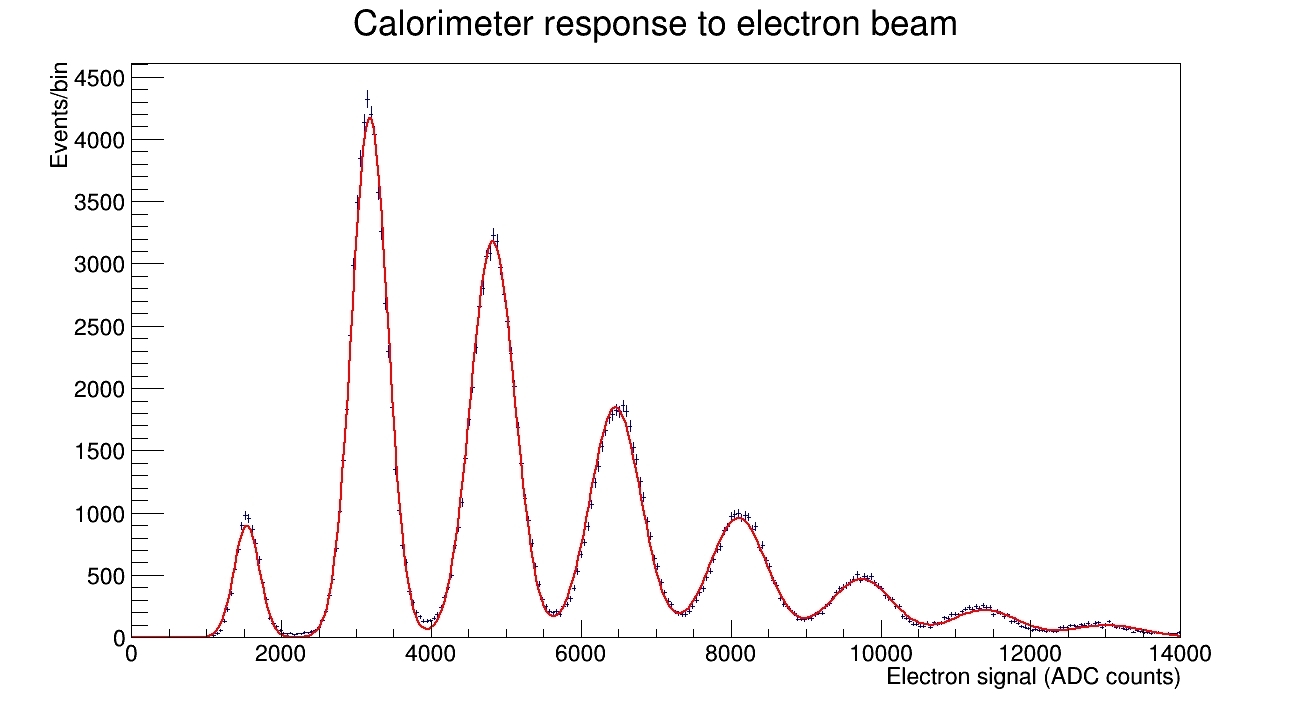}
\caption{Calorimeter response showing single and multiple electron peaks, together with fitted curve.}
\label{pe_spectrum}
\end{figure}

An example of the distribution of the calorimeter response is shown in fig.~\ref{pe_spectrum}; given the small number of electrons per spill the single and multiple 450 MeV electron peaks are clearly observable. We  fit this distribution with a sum of Gaussian distributions, where the means are assumed to be linearly related to the number of electrons and the widths with their square root. The assumption on the widths is based on the Poisson statistics of the number of fired pixels and on a contribution from the beam energy spread. The fit is typically well behaved and returns the mean value of the single electron peak in ADC counts. Dividing this value by the previously obtained calibrations and then by the beam energy we obtain the average number of fired pixels per MeV of about 0.9 pixels/MeV, consistent with expectations and previous measurements with the black crystal wrapping ~\cite{fienberg15}.
We can relate the laser pulse intensity to an equivalent electron energy by dividing the measured response with the unfiltered laser to the mean response for the single electron, and multiplying by the electron energy of 450 MeV. Typical values obtained in this test are around 800 MeV, which correspond to a measured light power before the filter wheel of 11.2 $\pm$ 1.1 pJ. This value can be scaled to the laser power predicted in the final full calorimeter system, where we expect 141 pJ before the filter wheel instead of the 11.2 pJ measured here. The equivalent maximum energy seen by each calorimeter cell would then be 800 MeV$\times$141 pJ/11.2 pJ $\simeq$ 10 GeV.
This calculation assumes an initial laser power of 1 nJ but, since the manufacturer of our laser heads guarantees a maximum power between 0.6 and 1.0 nJ, this prediction should be scaled with the maximum power available in the practice. In any case this light yield is well matched to the  3.1 GeV maximum electron energy expected in the calorimeter from muon decays in the muon $g$\,$-$\,$2$  experiment.

\subsection{Stability monitoring and corrections
\label{ssec:Y}}

The SiPM response to the electron incident on the calorimeter will be affected by the SiPM gain variations with temperature and bias voltage. Monitoring the response of the SiPMs to the laser pulses during data-taking allows tracking and correction of these variations. The response of a SiPM to the electron beam is given by
\begin{equation}
	r^{SiPM}_{el}(t)= R^{SiPM}_{el}\cdot f^{SiPM}_{gain}(t) 
,\end{equation}
where $R^{SiPM}_{el}$ is the SiPM response assuming a constant gain starting at time $t=0$ and $f^{SiPM}_{el}(t)$ is the time-dependent fluctuation in the SiPM's gain. The corresponding response to a laser pulse is:
\begin{equation}
r^{SiPM}_{laser}(t)= R^{SiPM}_{laser}(t)\cdot f^{SiPM}_{gain}(t) 
,\end{equation}
where $R^{SiPM}_{laser}(t)$ is the laser light received by the SiPM. The light received can in turn vary due to laser intensity and distribution chain fluctuations:
\begin{equation}
R^{SiPM}_{laser}(t)=R^{SiPM}_{laser}(t=0)\cdot f_{laser}(t)\cdot f_{distribution}(t) 
\end{equation}
where $f_{laser}(t)$ and $f_{distribution}(t)$ are determined respectively by the Source and Local Monitors. Corrected electron beam signal is then given by
\begin{equation}
\begin{aligned}
	&R^{SiPM}_{el}= \frac{r^{SiPM}_{el}(t)}{f^{SiPM}_{gain}(t)}\\
&= r^{SiPM}_{el}(t)\cdot
\frac{R^{SiPM}_{laser}(t=0) f_{laser}(t) f_{distribution}(t)}{r^{SiPM}_{laser}(t)}
\end{aligned}
\end{equation}

\begin{figure}[h!]	
 \begin{center}	
\includegraphics[width=0.5\textwidth]{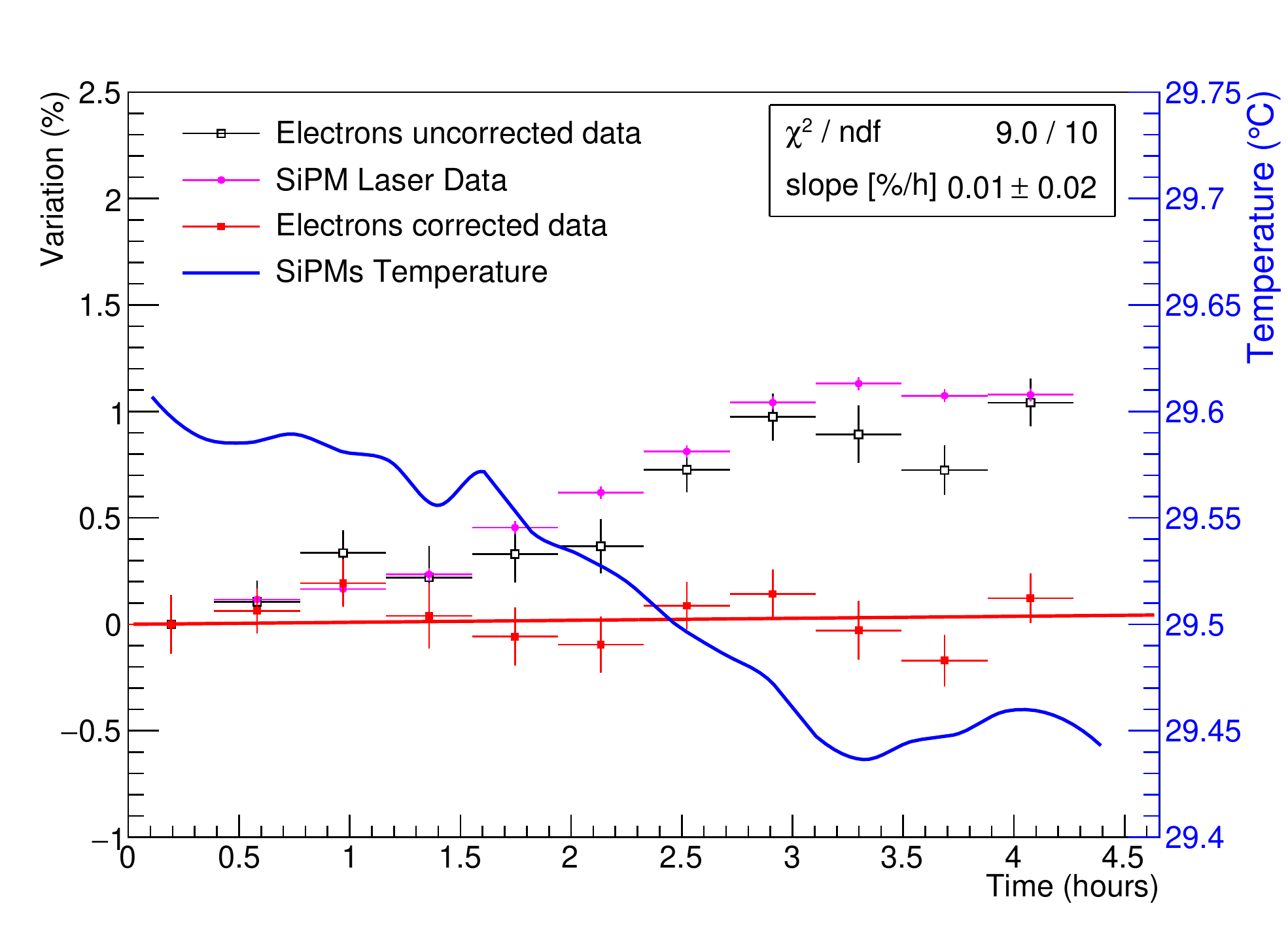}
	\end{center}	
	\caption{Variations in the measured energy of the electron beam and of the laser signals during four hours of data acquisition. The black (magenta) open circles show the gain fluctuations in the raw electron (laser) data while the full-red circles are the same data after the laser-based calibration correction has been applied. Variations are evaluated with respect to the first data points. The SiPM temperatures recorded during the same period are represented by the blue line.}
	\label{results}
\end{figure}

The result of this correction process is illustrated in figure~\ref{results} where the variations, relative to the first point, in the raw electron data taken during four hours of running are shown before and after correction.
Also shown in figure~\ref{results} are the variations in the laser data recorded during the same period. The corrected electron data correspond to the raw electron data divided by the corresponding laser data after correcting for laser intensity and light distribution stability.
Each point represents data averaged over approximately 23 minutes of running.

\begin{figure}[h!]	
 \begin{center}	
\includegraphics[width=0.5\textwidth]{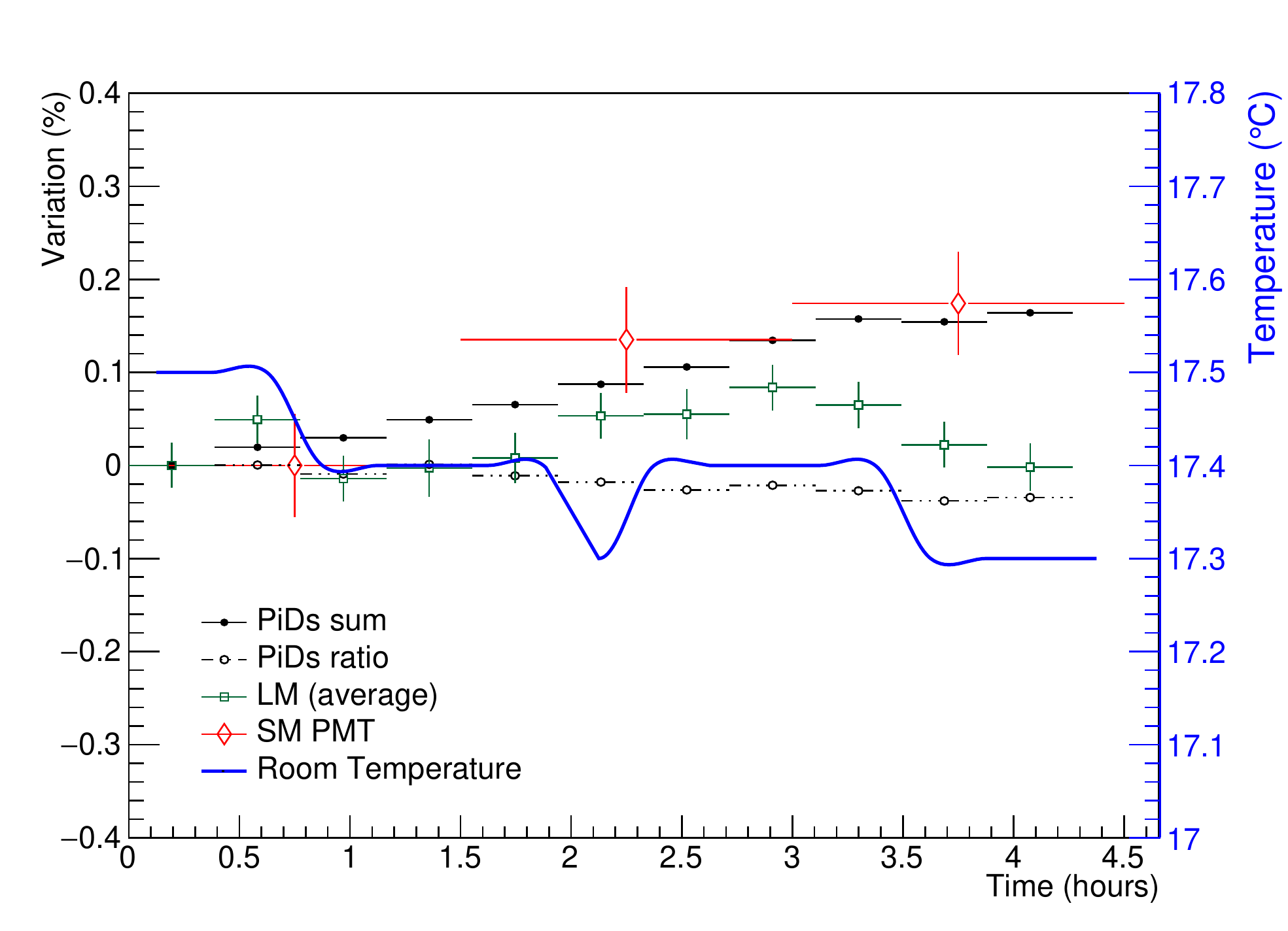}
	\end{center}	
	\caption{Stability of the laser calibration system. Solid black circles show the variations in the laser intensity as measured by the Source Monitor (SM) whereas the open squares represent the fluctuations in the laser light after distribution as recorded by the Local Monitor (LM). Variations are represented with respect to the first point of the distribution. The red diamonds are the corrected SM PMT data. The ratio between the two PiDs of the SM is also shown (black open squares). The ambient temperature recorded during the same period is shown by the blue line.}
	\label{corrections}
\end{figure}

\noindent As shown in Fig.~\ref{results}, the data without any corrections exhibit a positive drift of about 1.2\% over a four-hour run. The laser data are seen to track the electron data.
\noindent However, before using this laser data to correct for the SiPM gain variations, the laser data were, in turn, corrected for  fluctuations in the laser intensity and in the transmission efficiency of the laser pulses from the source to the calorimeter. These variations are shown in figure~\ref{corrections}, together with the ambient temperature recorded during the data-taking period (see next page).

\noindent The source monitor checks the stability of the laser intensity. The source monitor PiDs measured a variation of 0.2\%, as shown by the black solid circles in fig.~\ref{corrections}. Verification of this results can be obtained from the SM PMT which views the same laser pulses.
 All the fluctuations of the PMT response that do not depend on the laser fluctuations are corrected for by concurrently viewing the signals from the americium pulser located on the PMT's photocathode. These corrected PMT data are also shown as red diamonds in figure~\ref{corrections}. Given the very low activity of the Am-source incorporated in the pulser, a more accurate comparison requires longer periods of data-taking than were available during this test-beam run. Nevertheless, the data shown in figure~\ref{corrections} confirm the trend measured by the PiDs within their statistical accuracy. 
 
 The variations in the laser intensity, represented by the black points in figure~\ref{corrections}, correspond to variations in the average of the two PiDs viewing the same laser pulse in the SM. These data monitor the laser stability with a statistical precision of $0.003\%$ per point (23 minutes of data-taking in this case corresponding to 26,000 events). Given the large number of photoelectrons generated in each PiD  ($>10^{6}/$pulse) one would expect a statistical uncertainty  of  $<0.1\%$ per pulse or $0.0006\%$ for the 26,000 pulses collected. The much larger statistical error observed indicates that it is most likely driven by the noise of the prototype shaping amplifier used. This  may improve when the final version of this electronics will be used. Figure~\ref{corrections} also shows variations in the ratio of the two PiDs. This ratio is sensitive to variations in beam pointing which should cancel in the average.

Variations in the light transmission and distribution are measured by the LM. They correspond to the fluctuations in the ratio of the signal from the end of the optical transmission line to the signal monitored at the source. Since both signals are detected by the same PMT, and separated by about 100 ns, this ratio should be insensitive to fluctuations in the PMT gain. The mean of the fluctuations measured by the two local monitors is shown in figure~\ref{corrections} by open green squares. 



It is informative, at this point, to compare the variations of the SiPMs with the temperature variation measured by the sensors incorporated into the SiPM front-end electronics (blue line in figure~\ref{results}). It is clear, from this comparison that the SiPM gain variations are related to the temperature variations. 

Equivalent temperature measurements close to the PiDs and monitor PMTs are not available but, given the very low power dissipation of these devices and their front-end electronics, the ambient temperature of the experimental area is a good approximation. 

Temperature-related fluctuations in the photomultipliers of the LM are not relevant because (as previously explained) only the ratios of the nearly simultaneous signals from the same PMTs are necessary for the corrections. Since the ambient temperature variations (see figure~\ref{corrections}) are small, one does not expect a large effect on the PiD on the basis of the expected~\cite{hamamatsu} temperature dependence ($0.1\%/^{\circ}$C at 400 nm)~\cite{sipm}.  Temperature-dependence of the PiD response was nevertheless measured using a temperature controlled chamber which allowed control and measurement of the ambient conditions.

\begin{figure}[h!]	
 \begin{center}	
\includegraphics[width=0.5\textwidth]{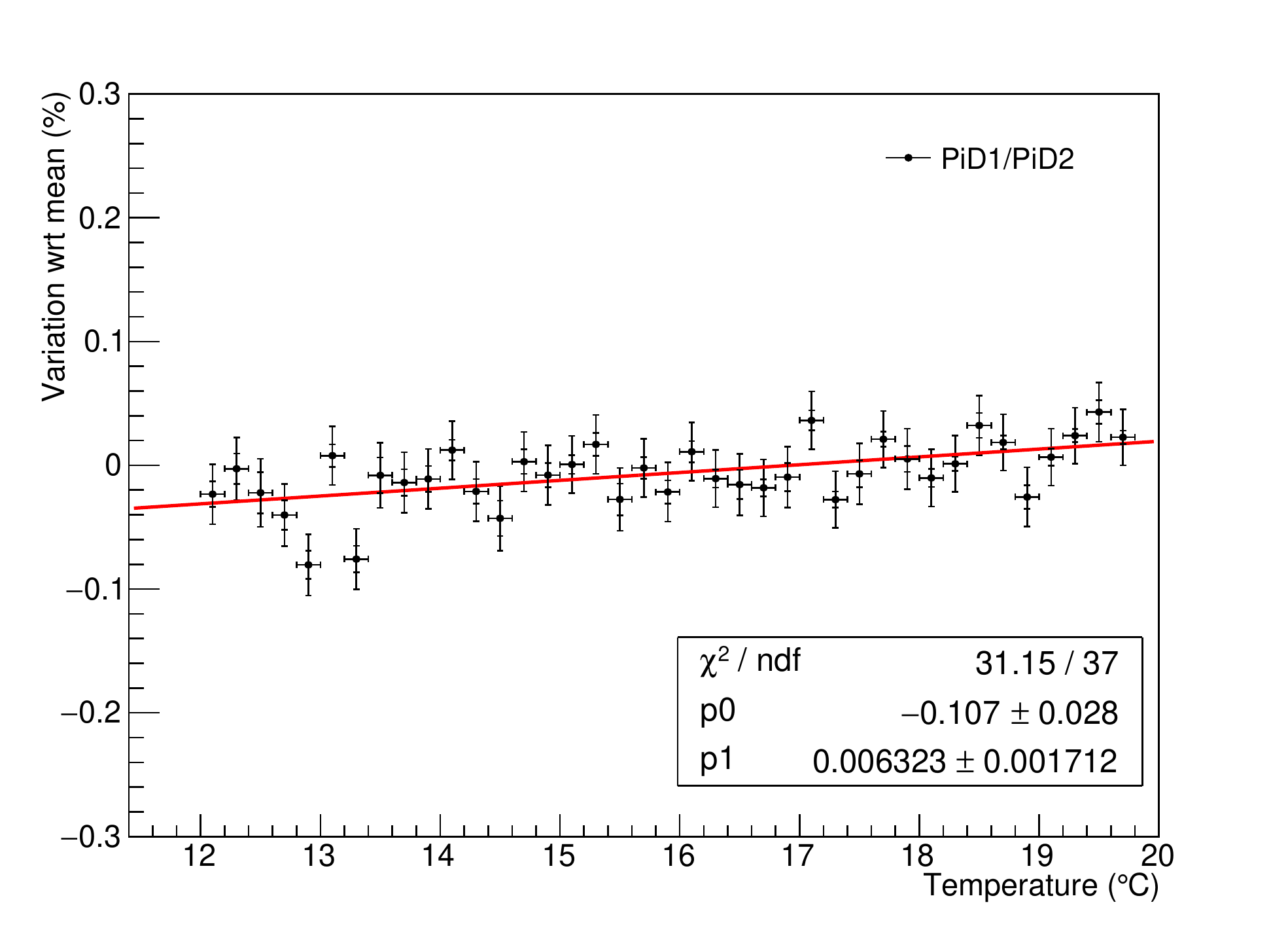}
	\end{center}	
	\caption{Temperature-dependence of the PiDs with frontend 
electronics measured at 450 nm. Error bars reflect the  statistical and 
total  (statistical and systematic) uncertainties.}
	\label{figura11}
\end{figure}

Measurements were made with a PiD inside the chamber (PiD1) and another identical one (the reference diode, PiD2) outside, both connected to their frontend electronics. The results of these measurements indicate that the response of the PiDs, coupled to their frontend electronics, is almost independent of temperature (see figure~\ref{figura11}).
These results indicate  a very good temperature stability of the PiDs and their electronics. An upper limit to the systematic error of 0.02 \% on the PiD response may be estimated by assuming the published temperature coefficient~\cite{hamamatsu} and a maximum temperature variation within $0.2\%/^{\circ}$C as achieved during more than four hours of our test. The variation of the PiD response reported in figure~\ref{corrections} reflects therefore true variations of the laser intensity.

The temperature-dependence of the NaI response reported in the literature~\cite{NaI} is comparable to that reported for the PiDs. 

\section{Conclusions}

Key elements of the full chain of the laser calibration system being developed for the $g$\,$-$\,$2$ experiment at Fermilab have been tested during a 450 MeV electron beam run at the Frascati Beam Test Facility.

The electron-energy equivalent of the laser intensity was measured and it was found that up to 10 GeV of equivalent energy could be delivered to every single calorimeter cell. This measurement allowed us to establish that six lasers will be sufficient to calibrate all the 24 calorimeters in the E989 experiment. It was also verified that the system is presently able to monitor and correct for laser intensity variations at the $10^{-4}$ level with less than 1000 laser pulses.  Variations in the distribution chain can be corrected by the LM at the same level on a longer  timescale. 

\vspace{1cm}
{\bf Acknowledgements}
\\
This research was supported by Istituto Nazionale di Fisica Nucleare (Italy) and by the EU Horizon 2020 Research and Innovation Programme under the Marie Sklodowska-Curie Grant Agreement No. 690835. The authors thank C.Di Giulio and L.Foggetta for support at BTF and L. K. Gibbons for a careful reading of the manuscript and useful suggestions.
\\
\\


\end{document}